# An efficient pipeline to compute patient-specific cerebral aneurysm wall tension

Mostafa Jamshidian[1*[0000-0002-5166-171X]], Benjamin Zwick[1[0000-0003-0184-1082]], Arosha S Dissanayake[2[0000-0003-0587-1529]], Adam Wittek[1[0000-0001-9780-8361]], Timothy J Phillips[3[0000-0001-9023-5986]], Stephen Honeybul[2[0000-0001-6841-8356]], Graeme J Hankey[4,5[0000-0002-6044-7328]], Karol Miller[1[0000-0002-6577-2082]]

[1] Intelligent Systems for Medicine Laboratory, The University of Western Australia, Perth, Western Australia, Australia

[2] Department of Neurosurgery, Sir Charles Gairdner Hospital, Perth, Western Australia, Australia

[3] Neurological Intervention and Imaging Service of Western Australia, Sir Charles Gairdner Hospital, Western Australia, Australia

[4] Centre for Neuromuscular and Neurological Diseases, Medical School, The University of Western Australia, Perth, Western Australia, Australia

[5] Perron Institute for Neurological and Translational Science, Perth, Western Australia, Australia

mostafa.jamshidian@uwa.edu.au

**Abstract.** Cerebral aneurysm rupture, leading to subarachnoid hemorrhage with a high mortality rate, disproportionately affects younger populations, resulting in a significant loss of productive life years. A significant proportion of these deaths is due to aneurysmal re-bleeding within the first three days following the initial bleed, prior to treatment. While early aneurysm treatment is recommended, there is no consensus on the ideal timing, and emergency treatment offers only an incremental benefit at a significant cost. Although various multivariable prediction models have been proposed to provide personalized risk assessments, no validated patient-specific predictor is available to rationalize emergency treatment. Furthermore, no model has yet incorporated emerging computational biomechanics-based biomarkers such as wall tension. In this paper, we proposed and validated an efficient semi-automatic pipeline to compute patient-specific cerebral aneurysm wall tension as a potential biomarker for the likelihood of re-bleeding. Our pipeline uses the patient's computed tomography angiography (CTA) image obtained at the time of subarachnoid hemorrhage diagnosis to create a patient-specific biomechanical model of the cerebral aneurysm using the finite element method. A distinctive feature of our approach is the straightforward model creation and wall tension computation using shell finite elements, without requiring patient-specific material properties or aneurysm wall thickness. Our non-invasive, patient-specific method for cerebral aneurysm wall tension can potentially provide individualized risk prediction and enhance clinical decision-making.

## 1 Introduction

Aneurysmal subarachnoid hemorrhage (aSAH), often due to the rupture of a saccular cerebral aneurysm [1], disproportionately affects a younger demographic [2], resulting in the loss of productive life years, which makes its impact nearly equivalent to that of the more common ischemic stroke [3]. One quarter of aSAH patients will die prior to reaching hospital [4], and up to one in six hospitalized patients will die prior to treatment within the first 72 hours, most often due to re-bleeding [5].

Whilst aneurysm treatment as soon as feasible is recommended, there is no consensus regarding the ideal timing due to the complex interplay between time-to-treatment and other risk factors, such as the degree of conscious state impairment and the thickness and distribution of the bleed. Routine emergency aneurysm treatment results in only a small incremental benefit but at a significant cost [6]. To rationalize emergency treatment, various multivariable prediction models have been proposed to provide personalized risk assessments, but they have not been systematically evaluated or externally validated. Additionally, these models cannot provide predictions for approximately 20% of patients with multiple causative aneurysms [7]. To date, no model has integrated emerging biomarkers such as aneurysm wall tension calculated by computational biomechanics [8] and early brain injury (EBI) markers [9, 10], although stress analysis of cerebral aneurysms has been researched previously [11-16].

In this study, we present an efficient semi-automatic software pipeline to compute patient-specific cerebral aneurysm wall tension as a possible biomarker for the likelihood of re-bleeding. In our pipeline, we use computed tomography angiography (CTA) images to create patient-specific biomechanical models of the cerebral aneurysms using finite element (FE) method, without requiring patient-specific material properties and aneurysm wall thickness. Specific to our approach is the implementation and verification of a straightforward approach for computing patient-specific wall tension in cerebral aneurysms using shell elements.

The remainder of the paper is organized as follows: In Section 2, we present the methods for wall tension computation, surface model creation, and biomechanical model creation using shell elements. We also describe the verification of the shell FE model against a solid FE model and the verification of our wall tension computation method. In Section 3, we present the results for the mesh convergence analysis, verification of the shell FE model, and verification of our approach for wall tension computation. Finally, Section 4 is devoted to the conclusions and discussions.

## 2 Methods

Based on mechanics, an artery ruptures when the local wall stress exceeds the local wall strength. Therefore, cerebral aneurysm wall stress, determined reliably and quickly through non-invasive approaches, can aid in patient-specific disease management. To compute cerebral aneurysm wall tension, we followed the approach implemented in the freely available BioPARR (Biomechanics-based Prediction of Aneurysm Rupture Risk) software package [17, 18]. Originally developed for computing wall stress in abdominal aortic aneurysms, BioPARR uses geometry from CTA images and mean arterial pressure as the applied load.

BioPARR stress recovery method, using linear FE analysis, is based on the observation that, as patient images demonstrate a deformed (due to blood pressure loading) artery geometry, the internal forces on the blood vessel wall that balance the applied pressure can be calculated based on statics principles [17, 19]. Accordingly, if the biomechanics simulations are arranged so that the aneurysm geometry remains unchanged when loaded by blood pressure, the stress field in the aneurysm wall balances the pressure [19]. Moreover, this stress field is shown to be independent of the material properties of artery tissue if the material is homogeneous in the current geometric configuration [20-24], although some studies have shown that this is also true for heterogeneous tissues [25].

In BioPARR, residual stresses are incorporated using Fung's Uniform Stress Hypothesis [26, 27], which involves averaging the stress across the artery wall thickness. This approach simplifies the computation of residual stresses, making it a straightforward post-processing step in aneurysm wall stress analysis [19]. We previously verified that, for a given material model and properties, the residual stress fields obtained using the BioPARR method are very close to those produced by the more complex non-linear iterative stress analysis method [26, 28].

## 2.1 Wall tension

Joldes et al. [26] proposed and validated a simple approach to compute arterial wall stress, incorporating residual stresses according to Fung's Uniform Stress Hypothesis [27]. The maximum principal wall stress $\bar{\sigma}$ based on the Uniform Stress Hypothesis is given by [26]:

$$\bar{\sigma} = \frac{1}{t}\int_0^t \sigma(\xi)\, d\xi, \tag{1}$$

where $\sigma(\xi)$ for $0 \leq \xi \leq t$ is the maximum principal stress (MPS) across the wall thickness $t$. Miller at al. [29] showed that maximum principal wall stress is proportional to the wall thickness. Therefore, inaccuracies in the difficult-to-measure wall thickness directly translate into errors in wall stress computation [30].

Considering the measurement difficulties, high uncertainties, and variations in patient-specific wall thickness of cerebral aneurysms, we used the maximum principal wall tension (MPWT) or simply wall tension, which is independent of wall thickness and is given by:

$$\bar{T} = \int_0^t \sigma(\xi)\, d\xi, \tag{2}$$

with units of force per length.

In FE analysis, MPWT is computed by numerical integration of MPS through thickness. Using an unstructured tetrahedral solid FE mesh, [26] computed the integral in Eq. (2) as a sum of piecewise integrals evaluated on several smaller subintervals of the wall thickness, using a two-point Gauss rule on each sub-interval. However, using a shell FE mesh, we have shown and verified in Section 3 that MPWT directly results from the outputs of the shell elements, with no numerical integration.

### 2.2 Image Segmentation and Surface Model

We used the 3D Slicer image computing platform [31, 32] to create the surface triangulation model of the cerebral aneurysm. We cropped a region of interest, encompassing the cerebral arteries with aneurysm, from Patient 1's fine-slice (voxel size 0.4~0.5 mm) CTA image obtained at the time of aSAH diagnosis, as shown in Fig. 1a. We semi-automatically segmented the geometry of cerebral arteries with aneurysm and created the surface model from the segmentation, as shown in Fig. 1b. Using the Vascular Modeling Toolkit (VMTK) [33] extension in 3D Slicer, we extracted the centerlines of vessels and clipped vessels to create the surface model of cerebral aneurysm with clipped connecting vessels, as shown in Fig. 1c. We saved this final surface triangulation model as an STL file and used it as the starting point for generating the FE mesh.

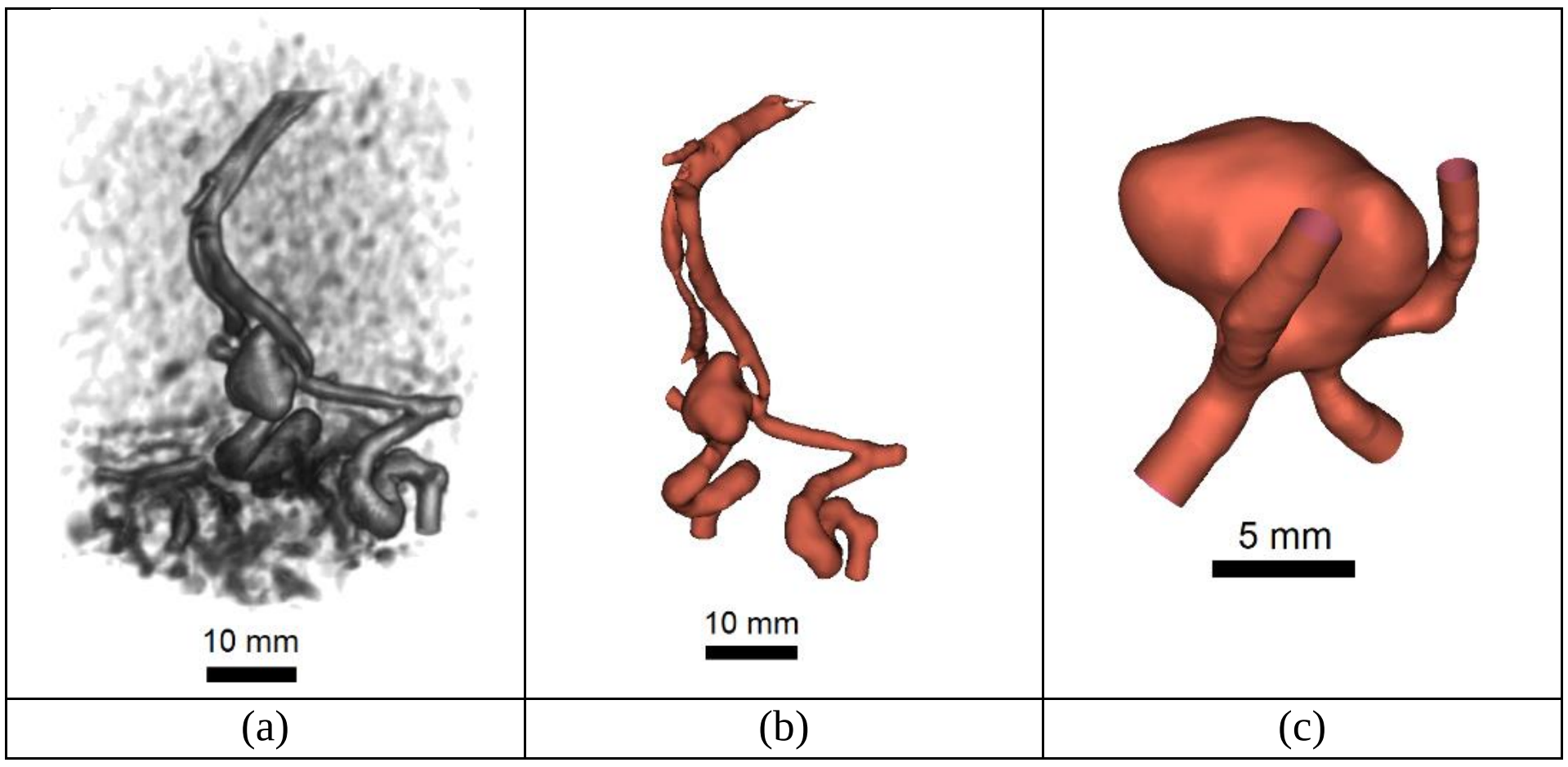

**Fig. 1.** Patient 1's cerebral aneurysm: (a) Volume rendering of the cropped 3D CTA image containing the cerebral arteries with aneurysm, (b) surface model of the cerebral arteries with aneurysm, (c) surface model of the cerebral aneurysm with clipped connecting vessels.

### 2.3 Biomechanical model

Using BioPARR methods, we computed the patient-specific wall tension of cerebral aneurysm through a direct linear analysis with no patient-specific material property data [19]. Following [11-13], we used shell elements in our FE analysis. Shell elements offer a significant computational advantage over solid elements for the analysis of thin-walled structures. By employing dimensional reduction, treating thickness as a parameter, and making simplified kinematic assumptions, shell elements require fewer elements and nodes. Aside from the simplicity of FE model creation, another major advantage of using shell elements for our application is the ease of computing wall tension from the output results of shell elements, as will be described later in Section 2.5.

To ensure the simplified kinematic assumptions do not affect the stress recovery results, we verified the stress recovery results of the shell FE analysis against those of a solid FE analysis for the same Patient 1's aneurysm geometry. Providing a full three-dimensional representation of the structure, solid elements can generally provide more accurate stress analysis for complex geometries and boundary conditions. We used the STL surface model of the aneurysm in Fig. 1c as the lumen boundary and created the aneurysm wall model by outwardly offsetting this surface. For the FE analysis, we used the Abaqus FE software [34].

**Shell FE mesh.** We re-meshed the surface model of the aneurysm in Fig. 1c, using VMTK to create a high-quality triangular shell mesh with desired element size. For mesh convergence analysis, we created shell meshes using four different element sizes of 0.15, 0.20, 0.3, and 0.6 mm, as shown in Fig. 2. We imported the shell mesh into Abaqus via the built-in STL Import plug-in. We used the Abaqus STRI65 quadratic shell element that is a 6-node triangular thin shell, with five degrees of freedom per node. The five degrees of freedom per node for the Abaqus STRI65 shell element are translations in the local x (in-plane), y (in-plane), and z (out-of-plane) directions, and rotations about the local x and y axes.

The STRI65 element has three integration points in the element plane. To obtain stress variation through the shell thickness as output, we instructed Abaqus to perform section integration during analysis. Abaqus conducted numerical integration at a user-defined odd number of integration points through thickness to calculate the stresses independently at each integration point. These through-thickness integration points are located at the positions of the in-plane integration points and are evenly distributed along the normal to the shell element plane. With an odd number of integration points, the first and last integration points are located at the innermost and outermost positions along the thickness direction, and the middle integration point is at the middle of the thickness.

Seshaiyer et al. [14]. reported cerebral aneurysm wall thicknesses between 44 µm and 128 µm, with an average wall thickness of 86 µm. Since wall tension is independent of wall thickness, we used a uniform shell thickness parameter of 0.086 mm [11, 12], with outwards thickness offsetting from the element plane for all elements. The outwards thickness offsetting implies that the shell mesh, shown in Fig. 2, represents the internal surface of the thin-walled vessel.

**Solid FE mesh.** We used Gmsh [35, 36] to generate a high-quality solid mesh of wedge elements, with specified in-plane element size and desired number of element layers through the thickness, based on the STL surface model of aneurysm in Fig. 1c. We instructed Gmsh to thicken the solid mesh by outwards offsetting so that the STL surface model, serving as the basis for solid mesh generation, overlapped with the internal surface of the solid mesh. For mesh convergence analysis, we created solid meshes with four different refinements including, the in-plane element sizes of 0.15 mm and three element layer through thickness, the in-plane element sizes of 0.20 mm and two element layer through thickness, the in-plane element sizes of 0.30 mm and two element layer through thickness, and the in-plane element sizes of 0.60 mm and one element layer through thickness. Consistent with the shell mesh, the thickness of the solid mesh for all four meshes was 0.086 mm [11, 12]. It should be noted that, as shown in Fig. 2, the solid mesh explicitly models the thickness, whereas the thickness in the shell mesh is a parameter. The solid mesh was exported from Gmsh as an Abaqus input file and imported into Abaqus as an orphan solid mesh. We used the Abaqus C3D15 continuum element, that is a 15-node quadratic wedge (triangular prism).

We used a linear elastic material model with a Poisson's ratio of 0.49 for the vessel wall tissue [11, 37]. We used a large elastic modulus of 100 GPa in our linear FE analysis for stress recovery.

We fixed the clipped ends of the connecting vessels and applied a uniform pressure of 13.332 kPa (100 mmHg) on the inner surface of the aneurysm [11]. We solved the FE model of Patient 1's cerebral aneurysm using Abaqus/Standard solver [34].

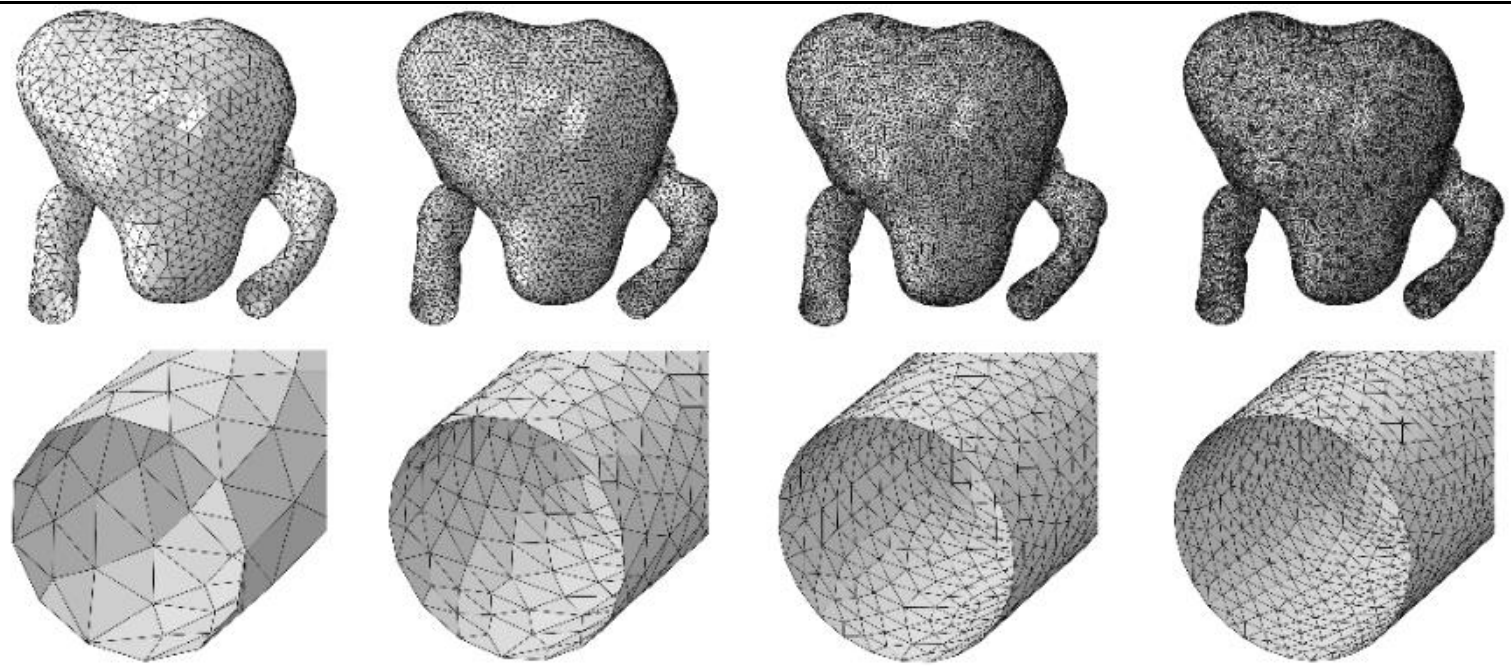

(a)

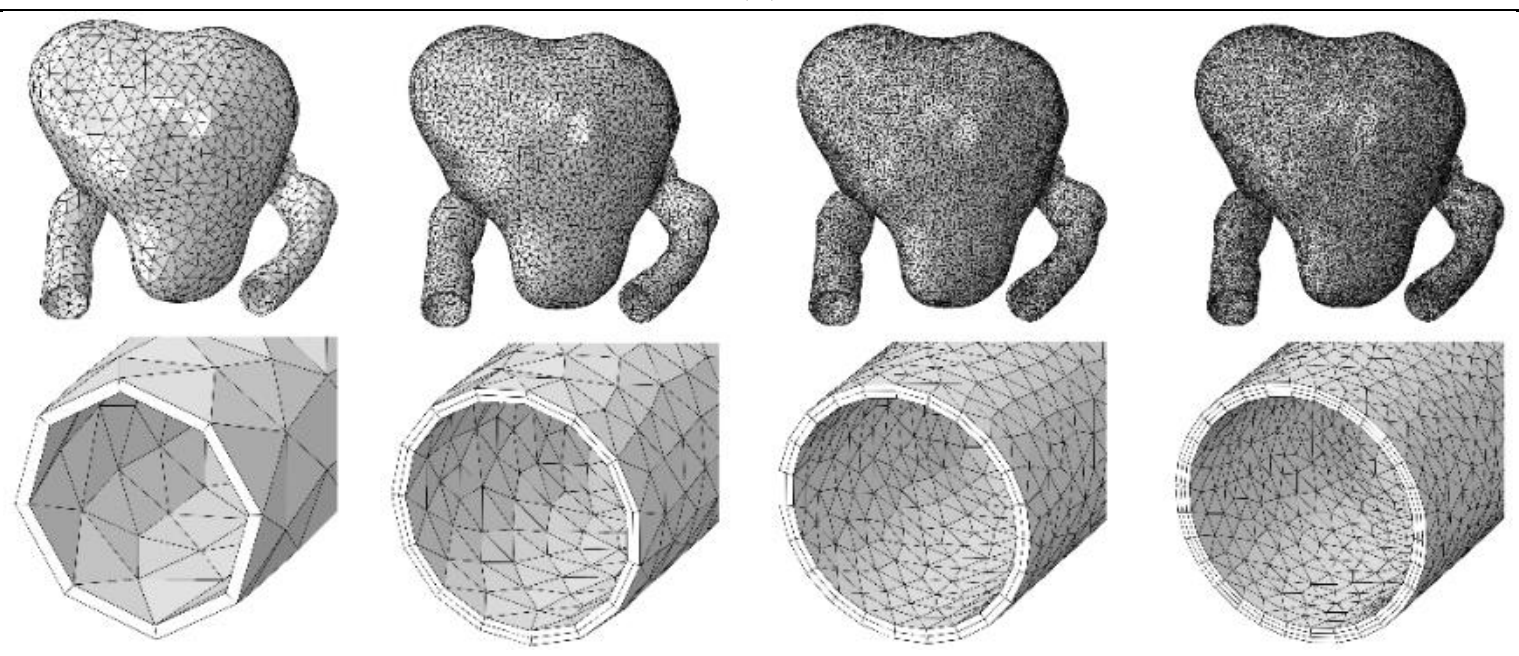

(b)

**Fig. 2.** (a) Shell and (b) solid finite element meshes used for mesh convergence analysis. Mesh refinement increases from left to right, with coarser meshes on the left and finer meshes on the right.

## 2.4 Method verification

For verification of mesh refinement via mesh convergence analysis, verification of shell mesh results against the solid mesh results, and verification of MPWT as a direct output of the shell FE analysis, we used stress percentile plots to compare two sets of stress results on the same domain (aneurysm wall) but represented on different sets of discrete points [38-40]. In the stress percentile plots, the horizontal axis indicates the percentile rank, and the vertical axis shows the corresponding percentile value.

**Verification of mesh refinement.** We performed the mesh convergence analysis for solid and shell meshes separately, each with four different mesh refinements, as shown in Fig. 2. For verification, we compared the percentile plots of the MPS instead of the MPWT because we did not implement the numerical computation of MPWT for the solid mesh.

**Verification of Shell mesh.** We verified the shell mesh FE analysis by comparing the percentile plots of MPS on the internal surface of the aneurysm between the shell and solid meshes, as they both originate from and share the same internal surface, as shown in Fig. 1c.

For the shell mesh, we created the MPS percentile plot using the MPS values at the innermost through-thickness integration points, as these points represent the internal surface of the aneurysm. However, in the solid mesh, the integration points lie within the thickness. We instructed Abaqus to report the MPS values on the internal surface of the solid mesh by interpolation using element shape functions. For the solid mesh, we created the MPS percentile plot using the MPS values at the interior FE nodes representing the internal surface.

It should be noted that in Abaqus, shell element outputs include both the maximum principal stress and the maximum in-plane principal stress. In this paper, for shell elements, MPS stands for the maximum in-plane principal stress output in Abaqus.

**Verification of wall tension:** For the numerical computation of MPWT for shell elements using Eq. (2), we instructed Abaqus to output the stresses at all through-thickness integration points. Using Eq. (2), we computed MPWT via numerical integration as follows:

$$\bar{T} = \frac{t}{n}\sum_{i=1}^{n}\sigma(i), \tag{3}$$

where $\sigma(i)$ is the MPS at the $i$th through-thickness integration point, with $i = 1, \dots, n$. For precise computation of MPWT, we used $n = 15$ through-thickness integration points.

It should be noted that the shell section forces, as a direct output from Abaqus, cannot be used to compute MPWT because these forces are provided in a local coordinate system that does not align with the MPS direction. According to Eq. (2), MPWT is the shell section force in the MPS direction.

Instead of numerical integration using Eq. (3), we have shown and verified in Section 3 that, in a linear analysis using a shell mesh, MPWT equals the MPS at the mid-surface through the thickness multiplied by the shell thickness. The MPWT field out-put was created in Abaqus by multiplying the shell thickness by the MPS field output at the mid through-thickness integration points, which is a direct output from Abaqus.

To verify this technique, we compared the percentile plots of MPWT computed via numerical integration using Eq. (3) with MPWT computed as MPS at mid through-thickness integration points multiplied by the shell thickness.

## 3 Results

Fig. 3 and Fig. 4 show the mesh convergence analysis results for solid and shell meshes, respectively, using the FE meshes shown in Fig. 2. The MPS on the internal surface of the cerebral aneurysm was used as the variable of interest in the mesh convergence analysis. In each figure, the MPS percentile plots illustrate convergence across various mesh refinements. From left to right, the MPS contour plots correspond to increasingly finer meshes. Fig. 3 and Fig. 4 demonstrate that the solution converges at an FE size of 0.3 mm for both shell and solid FE meshes. To account for variations in aneurysm geometry, we used an FE size of 0.2 mm as the sufficient mesh refinement for both shell and solid FE models. The Shell mesh with FE size of 0.2 mm consists of 74292 nodes and 37078 elements of type STRI65. The solid mesh with FE size of 0.2 mm consists of 212565 nodes and 60601 elements of type C3D15. Additionally, a separate convergence analysis on the number of through-thickness integration points determined that five integration points through the shell thickness are sufficient. These results are not presented here for brevity.

The simulation times for shell and solid FE models are 30 seconds and 70 seconds, respectively, on a laptop workstation with a 12th Gen Intel® Core™ i9-12900H 2.50 GHz processor and 64.0 GB of RAM. Although the difference in simulation time might be negligible for practical applications, the shell FE model is advantageous due to its straightforwardness in modeling and wall tension extraction.

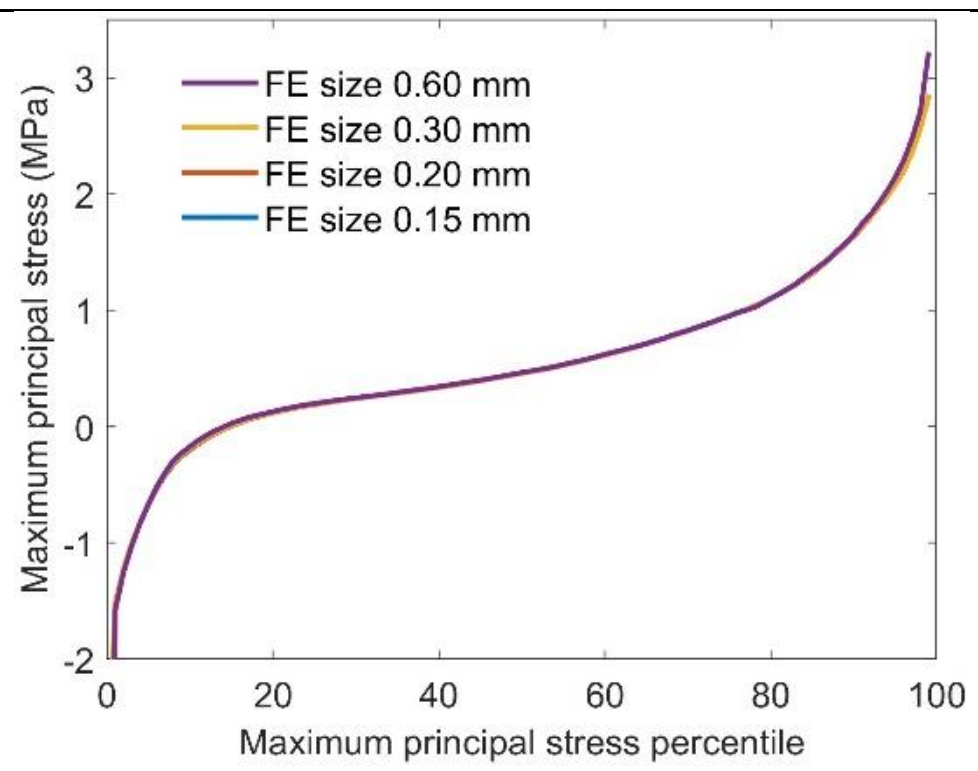

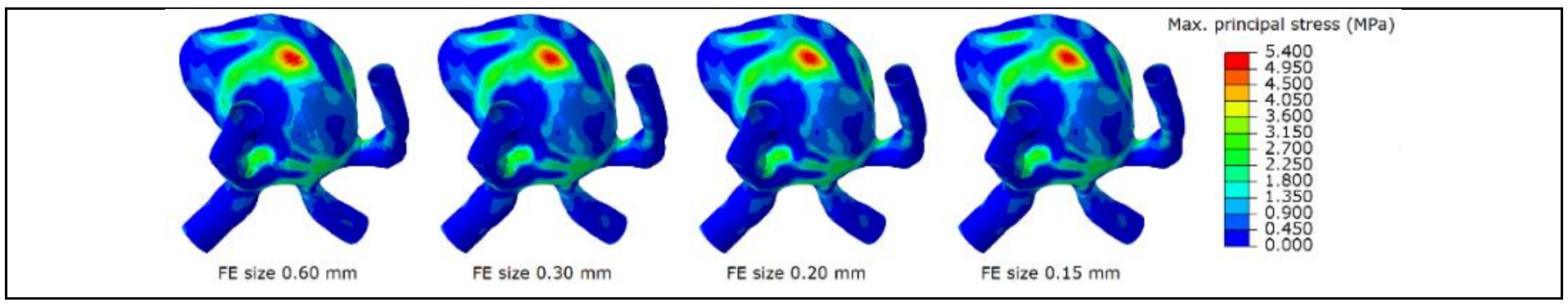

**Fig. 3.** Mesh convergence analysis for the shell finite element mesh using the maximum principal stress (MPS) on the internal surface of the cerebral aneurysm as the variable of interest. The MPS percentile plots show convergence across various mesh refinements. The MPS contour plots, from left to right, correspond to coarser to finer meshes.

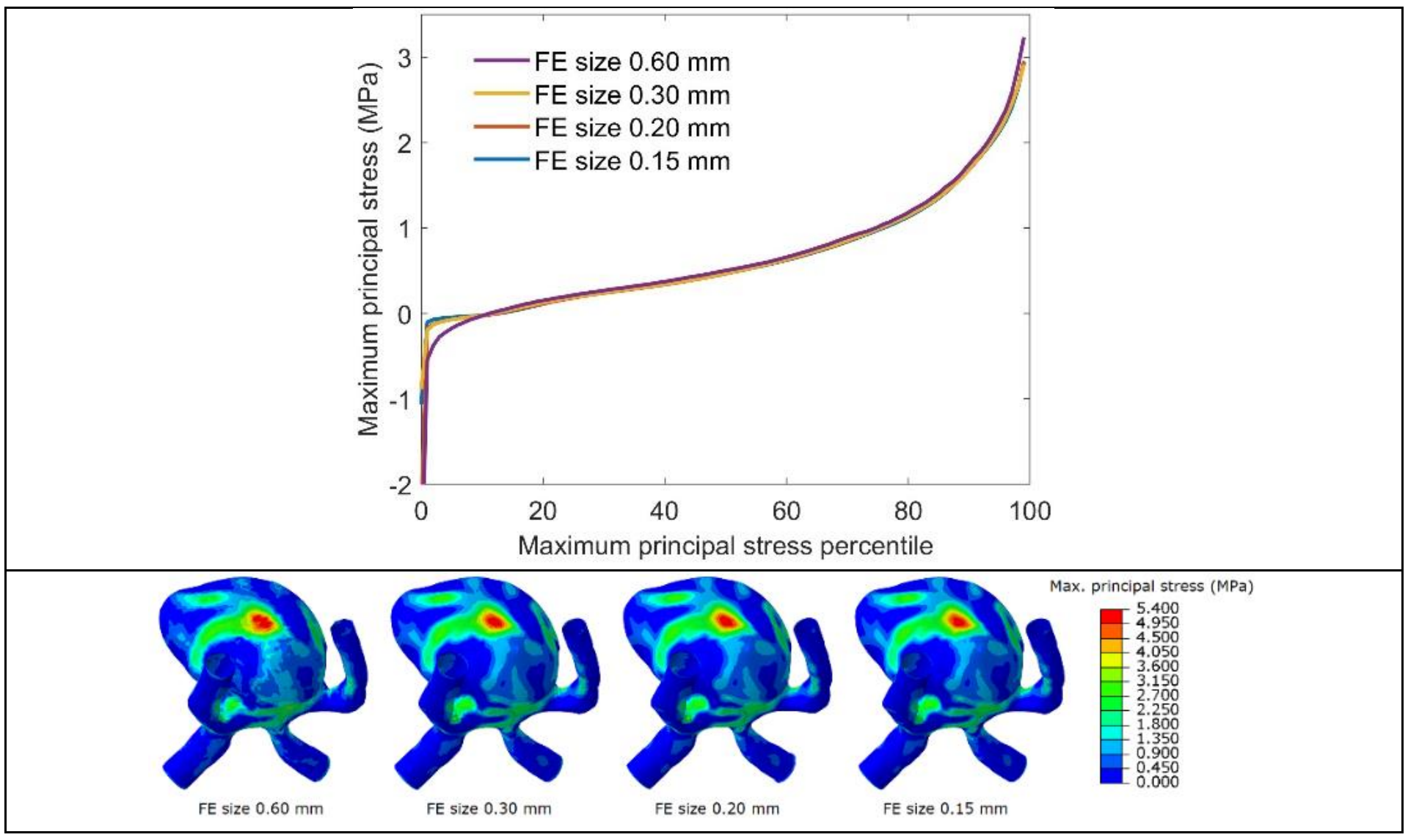

**Fig. 4.** Mesh convergence analysis for the solid finite element mesh using the maximum principal stress (MPS) on the internal surface of the cerebral aneurysm as the variable of interest. The MPS percentile plots show convergence across various mesh refinements. The MPS contour plots, from left to right, correspond to coarser to finer meshes.

Fig. 5 shows the verification of the shell FE model solution against the solid FE model solution by comparing their results for the MPS on the internal surface of the cerebral aneurysm shown in Fig. 1c. In addition to the MPS contour plots, Fig. 5 includes percentile plots of MPS, providing a detailed comparison between the two meshes. Fig. 5 demonstrates that the shell mesh produces almost identical MPS results to the solid mesh, except for MPS percentile ranks below the 15th.

Since the aim of our stress analysis is to assess rupture risk by comparing the maximum local wall stress against the local wall strength, our objective is to determine the maximum stress, and thus the discrepancies in MPS for low percentile ranks are not impacting our analysis. The differences in lower percentile ranks, corresponding to lower stress values, may be attributed to variations in MPS directions between the two models.

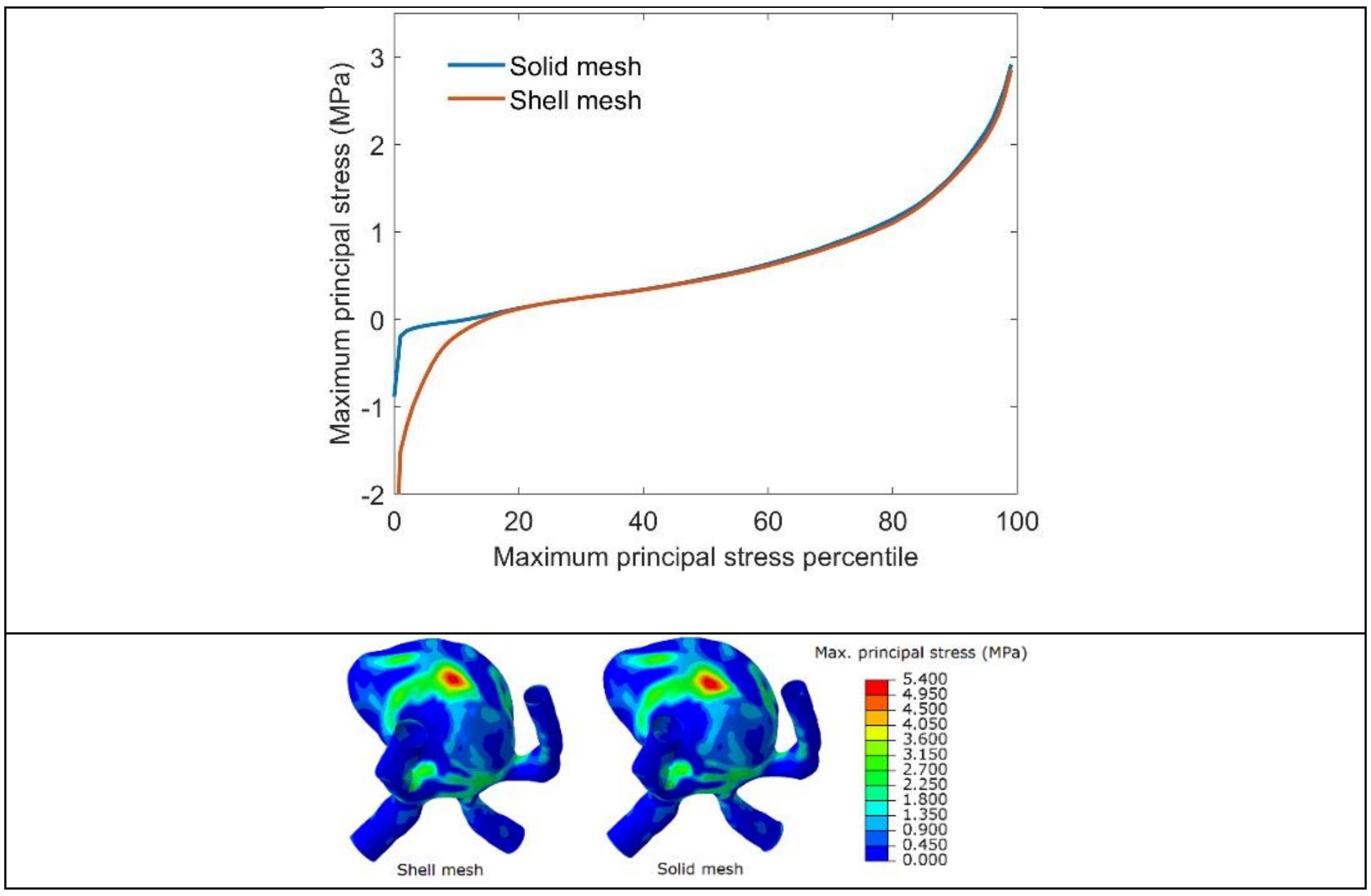

**Fig. 5.** Verification of the shell finite element mesh by comparing the maximum principal stress (MPS) on the internal surface of the cerebral aneurysm between the shell and solid meshes. MPS contour plots are presented alongside MPS percentile plots to provide a detailed comparison between the solid and shell meshes.

Fig. 6 demonstrates the validation of our straightforward method for computing wall tension. We validated our method by comparing MPWT computed as the MPS at the shell mid-surface multiplied by the shell thickness, against the MPWT obtained via numerical integration through the shell thickness. In Fig. 6, we compared the MPWT percentile plots obtained from both computation methods and, showing that our straightforward method for MPWT computation yields results nearly identical to those from the numerical integration method.

The contour plot in Fig. 6 shows the MPWT on Patient 1's cerebral aneurysm wall in N/mm, obtained by multiplying the shell thickness by the MPS at the mid-surface. As mentioned previously, in this paper, MPS for shell elements represents the maximum in-plane principal stress output in Abaqus. As shown in Fig. 6, for Patient 1, the maximum wall tension in the cerebral aneurysm is 0.197 N/mm.

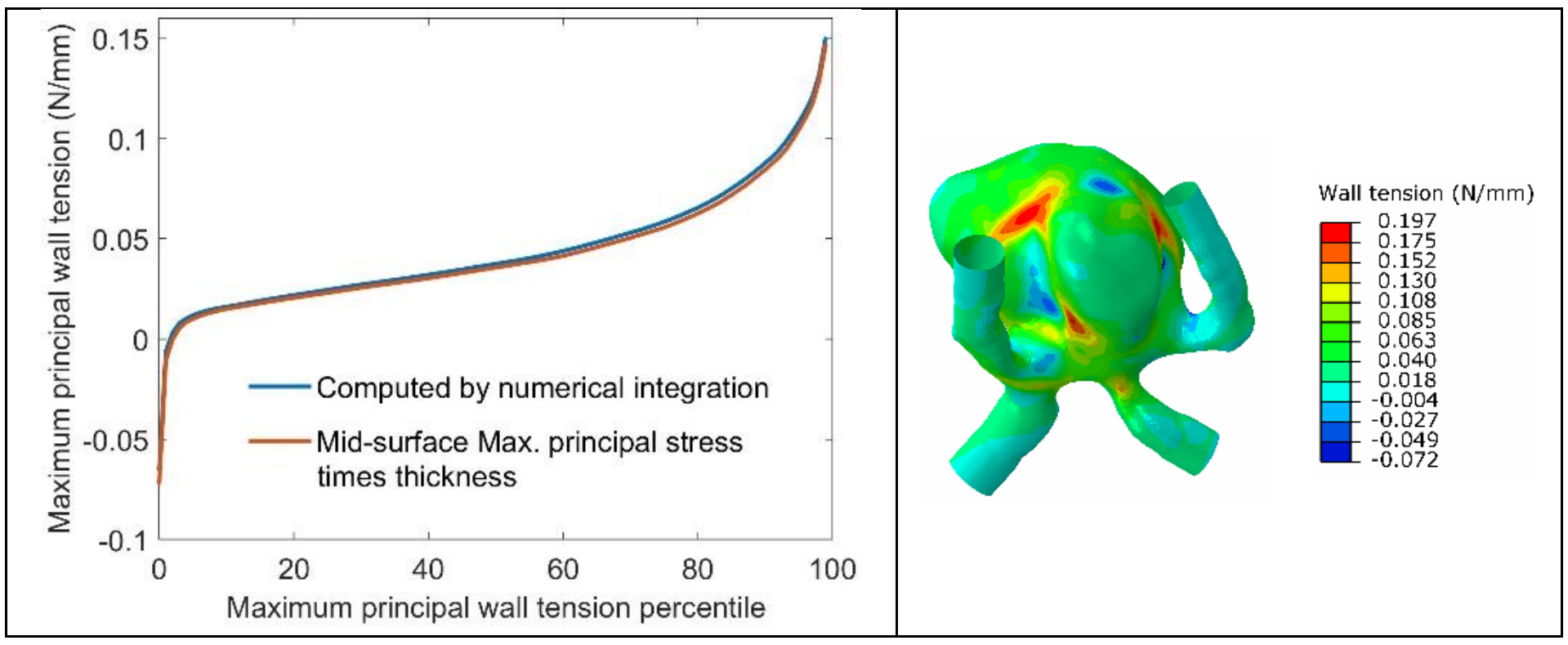

**Fig. 6.** Verification of the maximum principal wall tension (MPWT), computed as the maximum principal stress (MPS) at the shell mid-surface multiplied by the shell thickness, against MPWT computed via numerical integration through shell thickness. MPWT percentile plots obtained from both computation methods are compared. The contour plot illustrates MPWT on the cerebral aneurysm wall in N/mm obtained by multiplying the shell thickness by MPS at the mid-surface. Here, MPS represents the maximum in-plane principal stress output of shell elements in Abaqus.

To revisit and summarize, we employed our approach to compute cerebral aneurysm wall tension for another patient, referred to as Patient 2, as shown in Fig. 7. Fig. 7a shows the surface model of Patient 2's cerebral aneurysm with clipped connecting vessels, extracted from Patient 2's CTA image via image segmentation, centerline extraction, and vessel clipping. Fig. 7b shows the FE model using shell elements of size 0.2 mm. Fig. 7c shows the contour plot of the MPWT, with a maximum of 0.111 N/mm.

In practice, instead of beginning from scratch, we used a template Abaqus CAE file and updated the template FE model for every new geometry. Specifically, our efficient pipeline involves the following steps, with the surface model of the aneurysm as input: (1) open the existing template Abaqus CAE file, (2) import the new aneurysm geometry, (3) update the existing FE model with the new geometry, (4) update the node set for fixed boundary condition application, (5) update the interior surface for blood pressure application, (6) update the patient-specific blood pressure value, (7) solve the updated FE model using Abaqus/Standard, and (8) create the wall tension field output as thickness multiplied by the maximum in-plane principal stress at the mid-surface. Except for steps (4) and (5) that need user input, the other steps can be automated via Python programming in Abaqus. Since wall thickness has no effect on wall tension, the shell thickness parameter in the template FE model needs no update.

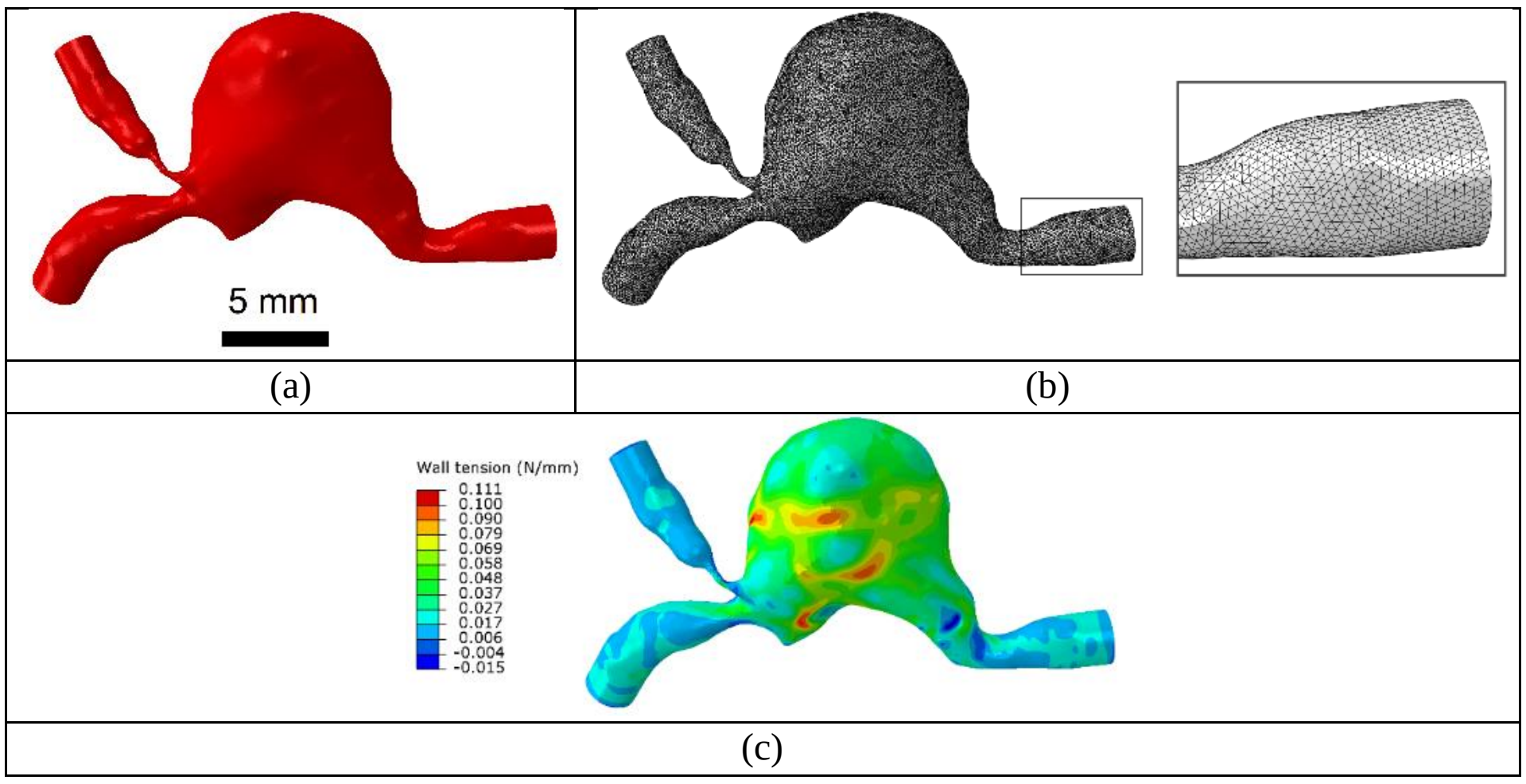

**Fig. 7.** Patient 2's cerebral aneurysm wall tension computed using the proposed efficient pipeline: (a) Aneurysm geometry, (b) Shell finite element mesh, (c) Contour plot of maximum principal wall tension.

## 4 Conclusions and Discussions

We presented an efficient streamlined approach for computing patient-specific cerebral aneurysm wall tension using finite element (FE) analysis of an aneurysm biomechanical model. The input to our pipeline is the CTA image of the aneurysm, and the output is the maximum principal wall tension (MPWT) map on the aneurysm wall. Our approach involves three major steps as follows: (1) semi-automatic image segmentation to extract the geometry of cerebral arteries with aneurysm, (2) creating the surface model of the cerebral aneurysm with clipped connecting vessels, and (3) FE analysis of the cerebral aneurysm biomechanical model to extract the MPWT map. Except for image segmentation in step (1), which requires an expert neurologist or radiologist, the other steps can be performed by an analyst with minimal training. Though steps (2) and (3) are not automated due to the variations and complexities in cerebral aneurysm geometry, they are straightforward and quick.

Our methodological innovation is the implementation and verification of an efficient approach for computing MPWT in cerebral aneurysms using shell elements. The shell FE model significantly simplifies the FE model creation and wall tension extraction processes. Our approach eliminates the need for numerical integration through the aneurysm wall thickness for computing MPWT. In our approach, MPWT is easily computed in Abaqus by multiplying the shell thickness by the maximum in-plane principal stress at the mid-surface, which is a direct output from Abaqus. In practice, the FE model creation is simplified by starting from a template Abaqus FE model and updating it for a new aneurysm geometry.

We assumed linear elastic, isotropic material properties for the cerebral aneurysm wall and a constant wall thickness in our analysis. However, these assumptions and simplifications do not compromise the validity of our results. In previous studies on abdominal aortic aneurysms (AAA), we demonstrated that, from the perspective of distinguishing between AAAs that would rupture and those that would remain intact, linear elastic analysis with isotropic material properties yields practically the same results as more complex non-linear models [39]. Furthermore, our simulations with varying lengths of the clipped connecting vessels (not shown here for brevity) indicate that the clipping location has a negligible effect on aneurysm wall tension.

Low image resolution and poor soft tissue contrast impair the accuracy and reliability of artery wall thickness measurements from CT images. As a result, constant wall thickness has been assumed in this and many other studies on cerebral aneurysm biomechanics [11, 12].

In conclusion, our approach allows for the non-invasive computation of patient-specific wall tension without requiring material properties and wall thickness. It can be used to investigate the correlation between aneurysm wall tension and the risk of aneurysmal re-bleeding. Once validated, the pipeline can be utilized in patient-specific disease management, potentially improving patient outcomes through more tailored treatment plans.